\documentclass[lettersize,journal]{IEEEtran}
\usepackage{amsmath,amsfonts}
\usepackage{cite}
\usepackage[numbers]{natbib}
\usepackage{notoccite}
\usepackage{algorithmic}
\usepackage{array}
\usepackage[caption=false,font=normalsize,labelfont=sf,textfont=sf]{subfig}
\usepackage{textcomp}
\usepackage{stfloats}
\usepackage{url}
\usepackage{verbatim}
\usepackage{graphicx}
\usepackage{hyperref}
\DeclareMathOperator*{\argmax}{argmax}
\def\BibTeX{{\rm B\kern-.05em{\sc i\kern-.025em b}\kern-.08em
    T\kern-.1667em\lower.7ex\hbox{E}\kern-.125emX}}
\usepackage{balance}
\usepackage{xcolor}

\begin{document}
\title{MISO Wireless Localization in The Presence of Reconfigurable Intelligent Surface}
\author{Sajjad Ghiasvand, Arash Nasri, Ali H. Abdollahi Bafghi, and Masoumeh Nasiri-Kenari,
	\IEEEmembership{Senior Member, IEEE}
\thanks{This work is supported by research affair center of Sharif University of Technology. The authors are with EE Dept., Sharif University of Technology. (\href{mailto:sajjad.ghiasvand@sharif.edu}{sajjad.ghiasvand@sharif.edu},\href{mailto:arash.nasri@ee.sharif.edu}{arash.nasri@ee.sharif.edu}, \href{mailto:aliabdolahi@ee.sharif.edu}{aliabdolahi@ee.sharif.edu}, \href{mailto:mnasiri@sharif.edu}{mnasiri@sharif.edu}).}}

\markboth{}%
{MISO Wireless Localization in The Presence of Reconfigurable Intelligent Surface}

\maketitle

\begin{abstract}
Reconfigurable Intelligent Surface (RIS) can play a pivotal role in enhancing communication rate and localization accuracy. In this letter, we propose a positioning algorithm in a RIS-assisted environment, where the Base Station (BS) is multi-antenna, and the Mobile Station (MS) is single-antenna. We show that our method can achieve a high-precision positioning if three RISs are available. We send several known signals to the receiver in different time slots and change the phase shifts of the RISs simultaneously in a proper way. Then, we propose a technique to eliminate the destructive effect of the angle-of-departure (AoD) in order to determine the distances between each RISs and the MS. Our numerical results indicate that the accuracy of the proposed algorithm is better than the algorithms that do not estimate the AoD.
\end{abstract}
\vspace{-0.35cm}
\section{Introduction}
\IEEEPARstart{O} 
ne of the most important technologies getting exceedingly popular is Reconfigurable Intelligent Surface (RIS)~\cite{RIS2}, which can be used for the enhancement of communication~\cite{communication} and the localization~\cite{localization}. RISs can be used as coatings for buildings and walls and can intelligently control the phase and amplitude of incident waves in the wireless channel. An essential part of these surfaces is a large number of elements, which can be used in passive or/and active modes~\cite{passiveOrActive}.\\
\indent
In the wireless localization problem, we aim to find the location of a Mobile Station (MS). Some of the pragmatic approaches for wireless localization are time of arrival (TOA)~\cite{TOA}, time differential of arrival (TDOA)~\cite{TOA}, received signal strength (RSS)~\cite{TOA}, angle of arrival (AOA)~\cite{TOA}, and  roundtrip time of flight (RTOF)~\cite{TOA} approaches. RISs can be used to achieve the position of the MS with higher precision~\cite{RISLocalization}. In this regard, authors in~\cite{RISLocalization},~\cite{linear},~\cite{obstructed} have proposed some methods in order to bound the localization error, but they have not proposed any practical method. In~\cite{nasri}, a novel algorithm for a high-precision localization has been proposed, where the transmitter and the receiver are single-antenna, and two (or more) RISs are available in the environment. Moreover, authors in~\cite{future} have explained some challenges, opportunities, and research directions that a RIS-assisted environment may experience.\\
\indent
In this letter, we consider RISs with passive elements and propose a localization algorithm to determine the location of the MS, where the Base Station (BS) is multi-antenna. We assume that there are three RISs in the environment, and we aim to find the distances between each RIS and the MS to determine its location. To this end, we use the method proposed in~\cite{nasri}. Our system model is, however, more generalized since we have considered the angle-of-departure (AoD) and the angle-of-arrival (AoA). In the proposed algorithm, the AoA and the AoD from the RISs to the MS, which are unknown, are eliminated. For AoD elimination, we send a known signal from the BS to the MS several times and change the phase shift of one of the RISs simultaneously while the phase shifts of other RISs are fixed. We can then find the strongest received signal at the MS to be able to calculate the distance between the RIS and the MS. Repeating this process for other RISs, we can obtain the distances between each RIS and the MS. \\
\indent
This letter is organized as follows. In Section II, we elaborate the system model. Then, in Section III, we introduce our localization algorithm and evaluate the performance of our method through some simulations in Section IV. Finally, the conclusion is presented in Section V.
\vspace{-0.25cm}
\section{System Model}
\begin{figure}[h]
 	\centering
 	\includegraphics[width=8cm]{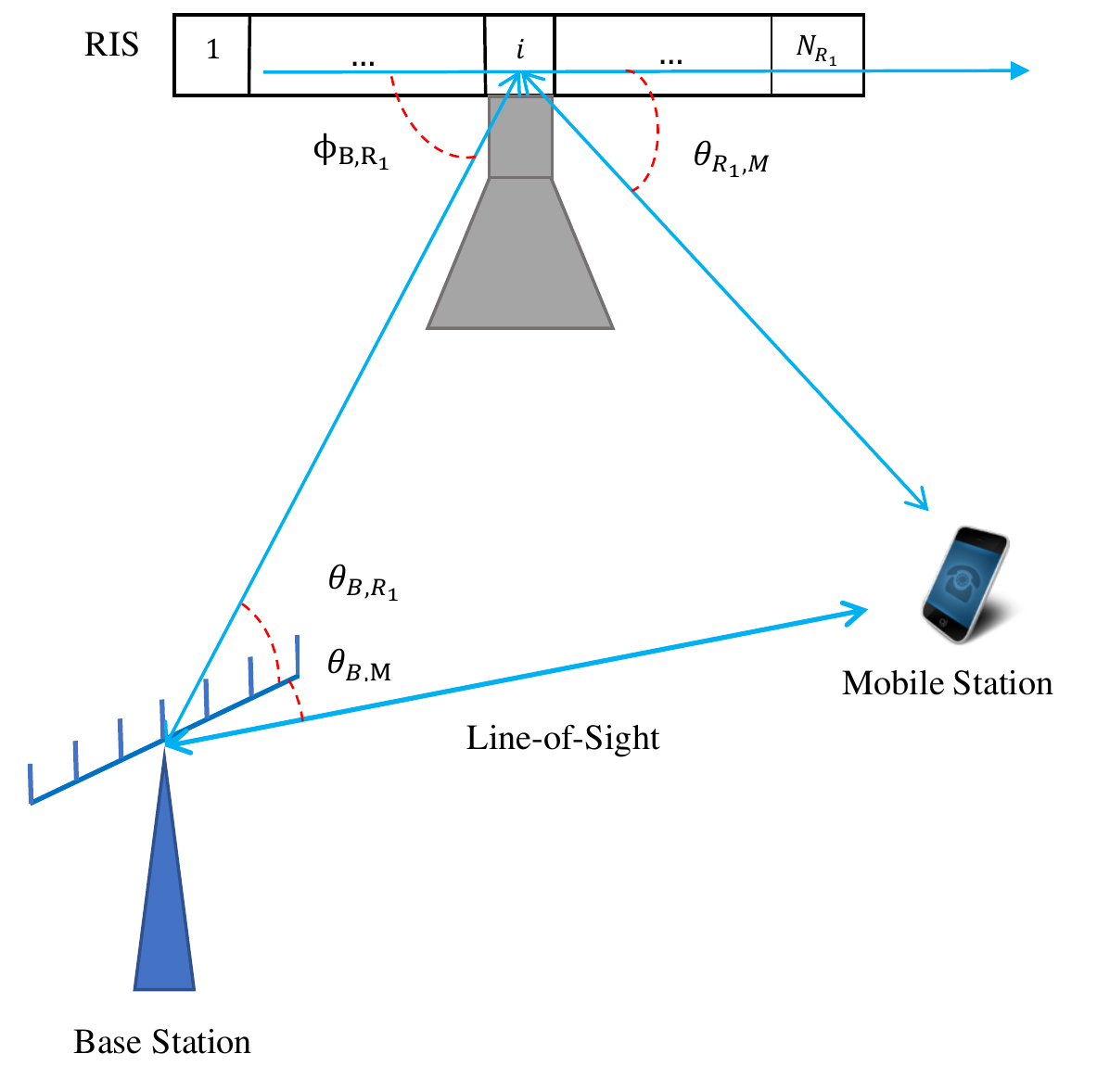}
 	\caption{The RIS-assisted environment, where the BS is equipped with $ N_B $ antennas, the MS is single-antenna, and three RISs are equipped with $ N_{R_1}, N_{R_2} $, and $ N_{R_3} $ elements (Only the first RIS has been shown in this figure). }
 	\label{fig0}
 \end{figure}

We consider a wireless localization system,
where the BS is equipped with $ N_B $ antennas, the MS is single-antenna, and three RISs are equipped with $ N_{R_\Gamma}$ elements for $ \Gamma \in \{1,2,3\} $. In this letter, we propose a localization algorithm for a three-dimensional scenario, where antennas and RISs are equipped with uniform linear arrays. We aim to determine the location of the MS while the location of the BS and RISs are given. Fig. \ref{fig0} illustrates the system model. We have considered the first RIS in this figure, and other RISs can be shown in a similar way. Based on this figure, $ \phi_{B,R_1} $ (AoA from the BS to the first RIS), $ \theta_{B,R_1} $ (AoD from the BS to the first RIS), $ \theta_{R_1,M} $ (AoD from the first RIS to the MS), and $ \theta_{B,M} $ (AoD from the BS to the MS) are defined in a three-dimensional space, so the model is also three-dimensional.
\noindent
We also use a channel model similar to the models used in~\cite{linear},~\cite{obstructed}. In our model there are four links: one direct link between the BS and the MS and three indirect links between the BS and the MS via three RISs. Each indirect link has two tandem channels, connecting the BS to the MS via one RIS.
The channels between the BS and each RIS are defined as:
\begin{equation}
\label{deqn_ex1}
\mathbf{H}_{B,R_\Gamma} = \rho_{B,R_\Gamma}e^{-j2 \pi f \tau_{B,R_\Gamma}} \boldsymbol{\alpha}_r(\phi_{B,R_\Gamma}) \boldsymbol{\alpha}_t^H(\theta_{B,R_\Gamma}),
\end{equation}
for $ \Gamma \in \{1,2,3\} $,
where $ \boldsymbol{\alpha}_r(\phi_{B,R_\Gamma}) \in \mathbb{C}^{N_{R_\Gamma}\times 1}$, whose $i$-th entry is $ [\boldsymbol{\alpha}_r(\phi_{B,R_\Gamma})]_i = e^{j(i-1)k \cos{\phi_{B,R_\Gamma}}} $, and $\boldsymbol{\alpha}_t(\theta_{B,R_\Gamma}) \in \mathbb{C}^{N_B\times 1} $, whose $i$-th entry is $ [\boldsymbol{\alpha}_t(\theta_{B,R_\Gamma})]_i = e^{j(i-1)k \cos{\theta_{B,R_\Gamma}}} $. $ k = 2\pi d/\lambda $, where $ d $ is the separation between BS's antennas, and $ \lambda $ is the wavelength of transmitted signal. $ \rho_{B, R_\Gamma} $ and $ \tau_{B, R_\Gamma} $ are free space path loss and phase shift due to the delay between BS and MS, and $ \theta_{B, R_\Gamma} $ and $ \phi_{B, R_\Gamma} $ are the  AoD and AoA, respectively. If we replace $ \boldsymbol{\alpha}_r(\phi_{B,R_\Gamma}) $ and $ \boldsymbol{\alpha}_t(\theta_{B,R_\Gamma}) $ in (\ref{deqn_ex1}) we have:
\begin{equation}
\label{deqn_ex2}
\small{
\begin{split}
\mathbf{H}_{B,R_\Gamma} = & \rho_{B,R_\Gamma}e^{-j2 \pi f \tau_{B,R_\Gamma}} 
\begin{bmatrix}
   e^{j(1-1)k \cos{\phi_{B,R_\Gamma}}} \\
  \vdots   \\
  e^{j(N_{R_\Gamma}-1)k \cos{\phi_{B,R_\Gamma}}}  \\
\end{bmatrix}
\\ &\times
\begin{bmatrix}
 e^{-j(1-1)k \cos{\theta_{B,R_\Gamma}}} & \cdots & e^{-j(N_B-1)k \cos{\theta_{B,R_\Gamma}}}\\
\end{bmatrix},
\end{split}
}
\end{equation}
where $ \mathbf{H}_{B, R_\Gamma} \in \mathbb{C}^{N_{R_\Gamma}\times N_B}$, whose entry in the $m$-th row and $n$-th column is $ [\mathbf{H}_{B, R_\Gamma}]_{m,n} = e^{jk [(m-1) \cos{\phi_{B,R_\Gamma}} - (n-1) \cos{\theta_{B,R_\Gamma}}]} $.\\
The channels between each RIS and MS are defined as:
\begin{equation}
\label{deqn_ex3}
\mathbf{H}_{R_\Gamma,M} = \rho_{R_\Gamma,M}e^{-j2\pi f \tau_{R_\Gamma,M}} \boldsymbol{\alpha}_r(\phi_{R_\Gamma,M}) 
\boldsymbol{\alpha}_t^H(\theta_{R_\Gamma,M}),
\end{equation}
for $ \Gamma \in \{1,2,3\} $,
where $ \boldsymbol{\alpha}_r(\phi_{R_\Gamma,M}) \in \mathbb{C} $ and $ \boldsymbol{\alpha}_t(\theta_{R_\Gamma,M}) \in \mathbb{C}^{N_{R_\Gamma}\times 1} $, and  $ \boldsymbol{\alpha}_r(\phi_{R_\Gamma,M}) $, $ \boldsymbol{\alpha}_t(\theta_{R_\Gamma,M}) $, $ \rho_{R_\Gamma,M} $, and $ \tau_{R_\Gamma,M} $ are defined similar to those in (\ref{deqn_ex1}). Since the MS is single-antenna, $ \boldsymbol{\alpha}_r(\phi_{R_\Gamma,M}) = 1 $. As a result, by replacing 
$ \boldsymbol{\alpha}_t(\theta_{R_\Gamma,M}) $ in (\ref{deqn_ex3}), $ \mathbf{H}_{R_\Gamma,M} $ can be written as:
\begin{equation}
\small{
\label{deqn_ex4}
\begin{split}
\mathbf{H}_{R_\Gamma,M} = &\rho_{R_\Gamma,M}e^{-j2 \pi f \tau_{R_\Gamma,M}}  \\ &\times
\begin{bmatrix}
 e^{-j(1-1)k \cos{\theta_{R_\Gamma,M}}} & \cdots & e^{-j(N_{R_\Gamma}-1)k \cos{\theta_{R_\Gamma,M}}}\\
\end{bmatrix},
\end{split}
}
\end{equation}
where $ \mathbf{H}_{R_\Gamma, M} \in \mathbb{C}^{1 \times N_{R_\Gamma}} $.\\
Finally, the channel between the BS and the MS is defined as:
\begin{equation}
\label{deqn_ex5}
\mathbf{H}_{B,M} = \rho_{B,M}e^{-j2 \pi f \tau_{B,M}} \boldsymbol{\alpha}_r(\phi_{B,M}) \boldsymbol{\alpha}_t^H(\theta_{B,M}),
\end{equation}
where $ \boldsymbol{\alpha}_r(\phi_{B,M}) \in \mathbb{C} $ and $ \boldsymbol{\alpha}_t(\theta_{B,M}) \in \mathbb{C}^{N_{B}\times 1} $, and  $ \boldsymbol{\alpha}_r(\phi_{B,M}) $, $ \boldsymbol{\alpha}_t(\theta_{B,M}) $, $ \rho_{B,M} $, and $ \tau_{B,M} $ are defined similar to those in (\ref{deqn_ex1}). Since the MS is single-antenna, $ \boldsymbol{\alpha}_r(\phi_{B,M}) = 1 $. As a result, by replacing 
$ \boldsymbol{\alpha}_t(\theta_{B,M}) $ in (\ref{deqn_ex5}), $ \mathbf{H}_{B,M} $ can be written as:
\begin{equation}
\small{
\label{deqn_ex6}
\begin{split}
\mathbf{H}_{B,M} = &\rho_{B,M}e^{-j2 \pi f \tau_{B,M}}  \\ &\times
\begin{bmatrix}
 e^{-j(1-1)k \cos{\theta_{B,M}}} & \cdots & e^{-j(N_{B}-1)k \cos{\theta_{B,M}}}\\
\end{bmatrix},
\end{split}
}
\end{equation}
where $ \mathbf{H}_{B, M} \in \mathbb{C}^{1 \times N_{B}} $.\\
We note that $ \rho_{A, B} $ and $ \tau_{A, B} $ can be obtained as:
\begin{equation}
\label{deqn_ex7}
\rho_{E, F} = || \mathbf{e} - \mathbf{f} ||_2^{-\mu/2} = \mathrm{\Delta R}^{-\mu/2}_{E, F}, \tau_{E, F} = \frac{|| \mathbf{e} - \mathbf{f} ||_2}{c},
\end{equation}
for $ E, F \in \{B, M, R_1, R_2, R_3\} $, where $ \mathrm{\Delta R}_{E, F} $ is the distance between nodes $ E $ and $ F $, $ \mu $ is the path loss exponent, and $ c $ is the speed of light.\\
The channel between the BS and the MS via each RIS can be then formulated as:
\begin{equation}
\label{deqn_ex8}
\mathbf{H}_\Gamma(\boldsymbol{\omega}_\Gamma) = \mathbf{H}_{R_\Gamma, M}\mathbf{\Omega}_\Gamma(\boldsymbol{\omega}_{\Gamma}) \mathbf{H}_{B, R_\Gamma}, \;\; \Gamma \in \{1,2,3\},
\end{equation}
where $ \boldsymbol{\omega}_\Gamma = [\omega_{\Gamma}^{(1)}, \cdots, \omega_\Gamma^{(N_{R_\Gamma})}]^T $,  $ \mathbf{H}_\Gamma(\mathbf{\boldsymbol{\omega}}_\Gamma) \in \mathbb{C}^{1 \times N_B} $, and $ \mathbf{\Omega}_\Gamma(\boldsymbol{\omega}_{\Gamma}) \in \mathbb{C}^{N_{R_\Gamma} \times N_{R_\Gamma}} $ is the phase shift control matrix at $\Gamma$-th RIS, defined as:
\begin{equation}
\label{deqn_ex9}
\mathbf{\Omega}_\Gamma(\boldsymbol{\omega}_{\Gamma}) =  \mathrm{diag}(\mathrm{exp}\{j\mathcal{\omega}_{\Gamma}^{(1)}\},\cdots, \mathrm{exp}\{j\mathcal{\omega}_{\Gamma}^{(N_{R_{\Gamma}})}\}).
\end{equation}
If we replace $ \mathbf{H}_{B, R_\Gamma} $ from (\ref{deqn_ex2}), $ \mathbf{H}_{R_\Gamma, M} $ from (\ref{deqn_ex4}), and $ \mathbf{\Omega}_\Gamma(\boldsymbol{\omega}_{\Gamma}) $ from (\ref{deqn_ex9}) in (\ref{deqn_ex8}), the $i$-th entry of $ \mathbf{H}_\Gamma(\boldsymbol{\omega}_{\Gamma}) $ is as follows:
\begin{equation}
\label{deqn_ex10}
\begin{split}
&[\mathbf{H}_\Gamma(\boldsymbol{\omega}_\Gamma)]_i = \\&\rho_{B,R_\Gamma}\rho_{R_\Gamma,M}e^{-j2 \pi f \tau_{B,R_\Gamma}} e^{-j2 \pi f \tau_{R_\Gamma,M}} e^{-jk [ (i-1) \cos{\theta_{B,R_\Gamma}}]} \\ & \times
\sum_{t = 1}^{N_{R_\Gamma}} e^{j\omega_{\Gamma}^{(t)}} e^{jk [(t-1) \cos{\phi_{B,R_\Gamma}} - (t-1) \cos{\theta_{R_\Gamma,M}}]}.
\end{split}
\end{equation}
Since RISs are equipped with uniform linear arrays we assume $ \hat{\boldsymbol{\omega}}_{\Gamma} = [(1-1)\varrho_\Gamma, \cdots, (N_{R_\Gamma} - 1)\varrho_\Gamma]^T $, so the sigma in (\ref{deqn_ex10}) can be written as:
\begin{equation}
\label{deqn_ex11}
\begin{split}
\Lambda_{\Gamma}(\hat{\boldsymbol{\omega}}_{\Gamma}) &= \sum_{t = 1}^{N_{R_\Gamma}} e^{j(t-1)\varrho_{\Gamma}} e^{jk [(t-1) \cos{\phi_{B,R_\Gamma}} - (t-1) \cos{\theta_{R_\Gamma,M}}]} \\&= \dfrac{1 - e^{jN_{R_\Gamma}[k(\cos\phi_{B,R_\Gamma} - \cos\theta_{R_\Gamma,M}) + \varrho_{\Gamma}]}}{1 - e^{j[k(\cos\phi_{B,R_\Gamma} - \cos\theta_{R_\Gamma,M}) + \varrho_{\Gamma}]}}.
\end{split}
\end{equation}
From (\ref{deqn_ex11}), we cannot determine $ \Lambda_{\Gamma}(\hat{\boldsymbol{\omega}}_{\Gamma}) $ since $ \theta_{R_\Gamma, M} $ is unknown; however, we can set the $ \hat{\boldsymbol{\omega}}_{\Gamma} $ as our desired value, and we will obtain $ \Lambda_{\Gamma}(\hat{\boldsymbol{\omega}}_{\Gamma}) $  in the next section. Moreover, we can rewrite the (\ref{deqn_ex10}) as follows:
\begin{equation}
\label{deqn_ex12}
\begin{split}
[\mathbf{H}_\Gamma(\hat{\boldsymbol{\omega}}_{\Gamma})]_i =& \rho_{B,R_\Gamma}\rho_{R_\Gamma,M}e^{-j2 \pi f \tau_{B,R_\Gamma}} e^{-j2 \pi f \tau_{R_\Gamma,M}} \\ & \times e^{jk [- (i-1) \cos{\theta_{B,R_\Gamma}}]} \Lambda_{\Gamma}(\hat{\boldsymbol{\omega}}_{\Gamma}).
\end{split}
\end{equation}
Finally, the entire channel between BS and MS can be formulated as:
\begin{equation}
\label{deqn_ex13}
\mathbf{H}(\boldsymbol{\omega}_1, \boldsymbol{\omega}_2, \boldsymbol{\omega}_3) = \mathbf{H}_{B,M} + \mathbf{H}_1(\boldsymbol{\omega}_{1}) + \mathbf{H}_2(\boldsymbol{\omega}_{2}) + \mathbf{H}_3(\boldsymbol{\omega}_{3}).
\end{equation}
We assume that no cross link exists between the RISs\footnote{In fact, due to the multiple reflections, the waves passing through cross-links become extremely weak and can be ignored.}.
Transmitter sends a known signal denoted by $ \mathbf{x} \in \mathbb{C}^{N_B \times 1} $, and the received signal can be expressed as:
\begin{equation}
\label{deqn_ex14}
\tilde{y}(\boldsymbol{\omega}_1, \boldsymbol{\omega}_2, \boldsymbol{\omega}_3) = \mathbf{H}(\boldsymbol{\omega}_1, \boldsymbol{\omega}_2, \boldsymbol{\omega}_3)\mathbf{x} + n,
\end{equation}
where $ n $ is a complex additive white Gaussian noise with variance $ \sigma^2 $.
\vspace{-0.25cm}
\section{Localization Algorithm}
In this section, first, the noise effect is ignored; then, we discuss the algorithm in the presence of noise in Remark 2 and consider its effects in our numerical results.
According to (\ref{deqn_ex14}), the received signal by the MS contains four terms. In~\cite{nasri}, a novel algorithm has been proposed to separate these terms when $ \boldsymbol{\omega}_1, \boldsymbol{\omega}_2, \boldsymbol{\omega}_3 \in \{\mathbf{0},\boldsymbol{\pi}\} $, where $ \mathbf{0}=[0, \cdots, 0]^T $, and $ \boldsymbol{\pi}=[\pi, \cdots, \pi]^T $. In this regard, we apply the proposed algorithm in~\cite{nasri} to our system model.\\
First, we send a known signal $ \mathbf{x} \in \mathbb{C}^{N_B \times 1} $ four times. At each transmission time, we choose the phase shift of RIS matrices $ \mathbf{\boldsymbol{\omega}}_1, \mathbf{\boldsymbol{\omega}}_2 $, and $ \mathbf{\boldsymbol{\omega}}_3 $ to separate the terms in (\ref{deqn_ex14}). At the first transmission slot, we select the phase shift of RIS matrices as $ \mathbf{\boldsymbol{\omega}}_1 = \mathbf{\boldsymbol{\omega}}_2 = \mathbf{\boldsymbol{\omega}}_3 = \mathbf{0} $, which leads to $ \boldsymbol{\Omega}_1 = \boldsymbol{\Omega}_2 = \boldsymbol{\Omega}_3 = \mathbf{I} $. Then, we have:
\begin{equation}
\label{deqn_ex15}
\tilde{y}_1 = (\mathbf{H}_{B,M} + \mathbf{H}_1(\mathbf{0}) + \mathbf{H}_2(\mathbf{0}) + \mathbf{H}_3(\mathbf{0}))\mathbf{x}.
\end{equation}
At the second transmission slot, we select phase shift of RIS matrices as $ \boldsymbol{\omega}_1 = \mathbf{\boldsymbol{\omega}}_3 = \mathbf{0}, \mathbf{\boldsymbol{\omega}}_2 = \boldsymbol{\pi} $, which leads to $ \boldsymbol{\Omega}_1 = -\boldsymbol{\Omega}_2 = \boldsymbol{\Omega}_3 = \mathbf{I} $. As a result the second received signal is:
\begin{equation}
\label{deqn_ex16}
\tilde{y}_2 = (\mathbf{H}_{B,M} + \mathbf{H}_1(\mathbf{0}) + \mathbf{H}_2(\mathbf{\pi}) + \mathbf{H}_3(\boldsymbol{0}))\mathbf{x}.
\end{equation}
At the third transmission slot, we select RIS matrices as $ \mathbf{\boldsymbol{\omega}}_3 = \mathbf{0}, \mathbf{\boldsymbol{\omega}}_1 = \mathbf{\boldsymbol{\omega}}_2 = \boldsymbol{\pi} $, which leads to  $ \boldsymbol{\Omega}_1 = \boldsymbol{\Omega}_2 = -\boldsymbol{\Omega}_3 = \mathbf{-I} $. Then, the third signal is:
\begin{equation}
\label{deqn_ex17}
\tilde{y}_3 = (\mathbf{H}_{B,M} + \mathbf{H}_1(\mathbf{\pi}) + \mathbf{H}_2(\boldsymbol{\pi}) + \mathbf{H}_3(\boldsymbol{0}))\mathbf{x}.
\end{equation}
Finally, we select RIS matrices as $ \mathbf{\boldsymbol{\omega}}_1 = \mathbf{\boldsymbol{\omega}}_2 = \mathbf{0},  \mathbf{\boldsymbol{\omega}}_3 = \boldsymbol{\pi} $, which leads to  $ \boldsymbol{\Omega}_1 = \boldsymbol{\Omega}_2 = -\boldsymbol{\Omega}_3 = \mathbf{I} $. Then, the forth signal is:
\begin{equation}
\label{deqn_ex18}
\tilde{y}_4 = (\mathbf{H}_{B,M} + \mathbf{H}_1(\mathbf{0}) + \mathbf{H}_2(\boldsymbol{0}) + \mathbf{H}_3(\boldsymbol{\pi}))\mathbf{x}.
\end{equation}
By noticing that $ \mathbf{H}_{\Gamma}(\mathbf{0}) = -\mathbf{H}_{\Gamma}(\boldsymbol{\pi}) $, and solving the four equations form (\ref{deqn_ex15}) to (\ref{deqn_ex18}), we obtain each term as follows:
\begin{equation}
\label{deqn_ex19}
\begin{split}
& y_1(\mathbf{0}) = \mathbf{H}_1(\mathbf{0}) \mathbf{x} = \frac{\tilde{y}_2 - \tilde{y}_3}{2},\;
y_2(\mathbf{0}) = \mathbf{H}_2(\mathbf{0}) \mathbf{x} = \frac{\tilde{y}_1 - \tilde{y}_2}{2},\;\\
& y_3(\mathbf{0}) = \mathbf{H}_3(\mathbf{0}) \mathbf{x} = \frac{\tilde{y}_4 - \tilde{y}_1}{2},\; 
y_0 = \mathbf{H}_{B,M}\mathbf{x} = \frac{\tilde{y}_3 + \tilde{y}_4}{2}.
\end{split}
\end{equation}
For simplicity, we defined:
\begin{equation}
\label{deqn_ex20}
y_\Gamma(\boldsymbol{\omega}_{\Gamma}) = \mathbf{H}_\Gamma(\boldsymbol{\omega}_{\Gamma}) \mathbf{x}, \; \Gamma \in \{1,2,3\}.
\end{equation}
As a result, the received signals by the MS when $ \mathbf{\boldsymbol{\omega}}_1 = \mathbf{\boldsymbol{\omega}}_2 = \mathbf{\boldsymbol{\omega}}_3 = \mathbf{0} $ are obtained from (\ref{deqn_ex19}).
Now, we consider $ \boldsymbol{\omega}_{1} = \boldsymbol{\omega}_{2} = \mathbf{0} $, and $ \boldsymbol{\omega}_{3} $ to be a arbitrary angle. Then, a known signal $ \mathbf{x} \in \mathbb{C}^{N_B \times 1} $ is sent to the MS. From (\ref{deqn_ex14}), (\ref{deqn_ex19}), and (\ref{deqn_ex20}) we have:
\begin{equation}
\label{deqn_ex21}
\tilde{y}(\mathbf{0}, \mathbf{0}, \boldsymbol{\omega}_{3}) = y_0 + y_{1}(\mathbf{0}) + y_{2}(\mathbf{0}) + y_{3}(\boldsymbol{\omega}_{3}).
\end{equation}
Therefore, $ y_{3}(\boldsymbol{\omega}_{3}) $ can be obtained as:
\begin{equation}
\label{deqn_ex22}
y_{3}(\boldsymbol{\omega}_{3}) = \tilde{y}(\mathbf{0}, \mathbf{0}, \boldsymbol{\omega}_{3}) - y_0 - y_{1}(\mathbf{0}) - y_{2}(\mathbf{0}),
\end{equation}
where $ \tilde{y} $ is the received signal, and $ y_{1}(\mathbf{0}) $, $ y_{2}(\mathbf{0}) $ are the ones obtained in (\ref{deqn_ex19}). We replace $ \mathbf{H}_\Gamma(\hat{\boldsymbol{\omega}}_{\Gamma}) $ from (\ref{deqn_ex12}) in (\ref{deqn_ex20}) to obtain $ y_{3}(\boldsymbol{\omega}_{3}) $ as:
\begin{equation}
\label{deqn_ex23}
\small{
\begin{split}
& y_{3}(\hat{\boldsymbol{\omega}}_{3}) = \rho_{B,R_{3}}\rho_{R_{3},M}e^{-j2 \pi f \tau_{B,R_{3}}} e^{-j2 \pi f \tau_{R_{3},M}} \Lambda_{3}(\hat{\boldsymbol{\omega}}_{3}) \\ & \times
 \begin{bmatrix}
                   e^{-jk [ (1-1) \cos{\theta_{B,R_{3}}}]},  & \cdots, & e^{-jk [ (N_B-1) \cos{\theta_{B,R_{3}}}]}\\ 
 \end{bmatrix}
\begin{bmatrix}
   x_1 \\
   x_2 \\
  \vdots   \\
   x_{N_B} \\
\end{bmatrix}\\
& =  \rho_{B,R_{3}}\rho_{R_{3},M} e^{j 2\pi f(\tau_{B,R_{3}} + \tau_{R_{3},M})} \cdot \Lambda_{3}(\hat{\boldsymbol{\omega}}_{3}) \\ & \times (e^{-jk [ (1-1) \cos{\theta_{B,R_{3}}}]}x_1 +  \ldots + e^{-jk [ (N_B-1) \cos{\theta_{B,R_{3}}}]}x_{N_B})\\
& =  \rho_{B,R_{3}}\rho_{R_{3},M} e^{j 2\pi f(\tau_{B,R_{3}} + \tau_{R_{3},M})} \Lambda_{3}(\hat{\boldsymbol{\omega}}_{3}) \cdot x \\ & \times \sum_{t = 1}^{N_B} e^{-j(t-1)k\cos(\theta_{B,R_{3}})}
\end{split}
}
\end{equation}
In the last equality, we have assumed that $ [\mathbf{x}]_i = x $. The value of the sigma in (\ref{deqn_ex23}) is known since $ \theta_{B, R_{3}} $ is known. We define $ \Xi_{3} = \sum_{t = 1}^{N_B} e^{-j(t-1)k\cos(\theta_{B,R_{3}})} $. Consequently $ y_{3}(\boldsymbol{\omega}_{3}) $ can be written as:
\begin{equation}
\label{deqn_ex24}
y_{3}(\hat{\boldsymbol{\omega}}_{3}) = \rho_{B,R_{3}}\rho_{R_{3},M} e^{j 2\pi f(\tau_{B,R_{3}} + \tau_{R_{3},M})} \Lambda_{3}(\hat{\boldsymbol{\omega}}_{3}) \cdot x \cdot \Xi_{3}.
\end{equation}
By calculating absolute value of both sides of (\ref{deqn_ex24}) we have:
\begin{equation}
\label{deqn_ex25}
|y_{3}(\hat{\boldsymbol{\omega}}_{3})| = \rho_{B,R_{3}}\rho_{R_{3},M} |\Lambda_{3}(\hat{\boldsymbol{\omega}}_{3}) \cdot x \cdot \Xi_{3}|.
\end{equation}
As a reminder $ \hat{\boldsymbol{\omega}}_{3} = [(1-1)\varrho_{3}, \cdots, (N_{R_\Gamma} - 1)\varrho_{3}]^T $. We sweep the phase shift of the third RIS, $ \varrho_{3} $, from $ 0 $ to $ 2\pi $, and send the known signal $ \mathbf{x} $ for each $ \varrho_{3} $. Then, the maximum value of $ |y_{3}(\hat{\boldsymbol{\omega}}_{3})| $ can be obtained for some $ \hat{\boldsymbol{\omega}}_{3} = \hat{\boldsymbol{\omega}}_{3}^* $ ($ \hat{\boldsymbol{\omega}}_{3}^* $ is not unique). Since by changing $ \hat{\boldsymbol{\omega}}_{3} $ just $ \Lambda_{3}(\hat{\boldsymbol{\omega}}_{3}) $ changes according to (\ref{deqn_ex24}), we have:
\begin{equation}
\label{deqn_ex26}
\hat{\boldsymbol{\omega}}_{3}^* = \argmax_{\hat{\boldsymbol{\omega}}_{3}} |\Lambda_{3}(\hat{\boldsymbol{\omega}}_{3})|.
\end{equation}
According to (\ref{deqn_ex11}), $ \Lambda_{3}(\hat{\boldsymbol{\omega}}_{3}) $ is a geometric series. Therefore, its maximum value happens when $ e^{j[k(\cos\phi_{B,R_{3}} - \cos\theta_{R_{3},M}) + \varrho_{3}]} = 1 $.
Therefore, for optimum values $ \hat{\boldsymbol{\omega}}_{3}^* $, we have:
\begin{equation}
\label{deqn_ex27}
\max_{\hat{\boldsymbol{\omega}}_{3}}|\Lambda_{3}(\hat{\boldsymbol{\omega}}_{3})| = |\Lambda_{3}(\hat{\boldsymbol{\omega}}_{3}^*)| = N_{R_{3}}.
\end{equation}
As a result, the maximum value of $ |y_{3}(\hat{\boldsymbol{\omega}}_{3})| $ in (\ref{deqn_ex28}) can be written as:
\begin{equation}
\label{deqn_ex28}
|y_{3}(\hat{\boldsymbol{\omega}}_{3}^*)| = \rho_{B,R_{3}}\rho_{R_{3},M} N_{R_{3}} |x \cdot \Xi_{3}|.
\end{equation}
$ y_{3}(\hat{\boldsymbol{\omega}}_{3}^*) $ is known by (\ref{deqn_ex22}).
Since $ \rho_{B,R_{3}} = \mathrm{\Delta R}^{-\mu/2}_{B,R_{3}} $ is known (the distances between the BS and RISs are given), we can obtain the distance between $\Gamma$-th RIS and MS as follows:
\begin{equation}
\label{deqn_ex29}
\Delta_{R_{3}, M} = \Big(\dfrac{|y_{3}(\hat{\boldsymbol{\omega}}_{3}^*)|}{\rho_{B,R_{3}}N_{R_{3}} | x \cdot \Xi_{3}|}\Big)^{-2/\mu}.
\end{equation}
In a similar way, we can find $ \hat{\boldsymbol{\omega}}_{1}^* $ and $ \hat{\boldsymbol{\omega}}_{2}^* $.
Thus, the distances between RISs and MS are known as follows:
\begin{equation}
\label{deqn_ex30}
\begin{split}
&\Delta_{R_1, M} = \Big(\dfrac{|\tilde{y}(\mathbf{0}, \mathbf{0}, \hat{\boldsymbol{\omega}}_{1}^*) - y_{2}(\mathbf{0}) - y_{3}(\mathbf{0}) - y_0|}{\rho_{B,R_1}N_{R_{1}} | x \cdot \Xi_{1}|}\Big)^{-2/\mu},\\
&\Delta_{R_2, M} = \Big(\dfrac{|\tilde{y}(\mathbf{0}, \mathbf{0}, \hat{\boldsymbol{\omega}}_{2}^*) - y_{1}(\mathbf{0}) - y_{3}(\mathbf{0}) - y_0|}{\rho_{B,R_2}N_{R_{2}} | x \cdot \Xi_{2}|}\Big)^{-2/\mu},\\
& \Delta_{R_3, M} = \Big(\dfrac{|\tilde{y}(\mathbf{0}, \mathbf{0}, \hat{\boldsymbol{\omega}}_{3}^*) - y_{1}(\mathbf{0}) - y_{2}(\mathbf{0}) - y_0|}{\rho_{B,R_3}N_{R_{3}} | x \cdot \Xi_{3}|}\Big)^{-2/\mu}.
\end{split}
\end{equation}
\textbf{Remark 1:} The proposed algorithm can be applied to RISs with rectangular arrays, which is almost similar to RISs with linear arrays. Nevertheless, because of the simplicity and the lack of space, we consider the RISs with linear arrays. In the scenario with rectangular arrays, there are two AoA and two AoD from the BS to the RISs, and there are two AoD from each RIS to the MS. To determine the distances between each RIS and the MS, the effect of two AoD from the RISs to the MS must be removed. To this end, each should be eliminated separately using the proposed algorithm. Other parts of this scenario are straightforward.\\
\textbf{Remark 2:} We could consider the noise and add it to the received signals in all aforementioned equations. For instance, in (\ref{deqn_ex24}), we would have:
\begin{equation}
\label{deqn_ex31}
y_{3}(\hat{\boldsymbol{\omega}}_{3}) = \rho_{B,R_{3}}\rho_{R_{3},M} e^{j 2\pi f(\tau_{B,R_{3}} + \tau_{R_{3},M})} \Lambda_{3}(\hat{\boldsymbol{\omega}}_{3}) \cdot x \cdot \Xi_{3} + n,
\end{equation}
where $ n $ is a complex Gaussian noise with the variance of $ \sigma^2 $. Then, we can consider real and imaginary parts of $ y_{3}(\hat{\boldsymbol{\omega}}_{3}) $ separately and have:
\begin{equation}
\label{deqn_ex32}
\begin{split}
& \underbrace{\mathrm{Re}(y_{3}(\hat{\boldsymbol{\omega}}_{3}))}_{y} = \\ & \underbrace{\mathrm{Re}(\rho_{B,R_{3}}\rho_{R_{3},M} e^{j 2\pi f(\tau_{B,R_{3}} + \tau_{R_{3},M})} \Lambda_{3}(\hat{\boldsymbol{\omega}}_{3}) \cdot x \cdot \Xi_{3})}_{\theta_{Real}} + \mathrm{Re}(n),
\end{split}
\end{equation}
\begin{equation}
\label{deqn_ex33}
\begin{split}
&\mathrm{Im}(y_{3}(\hat{\boldsymbol{\omega}}_{3})) = \\ & \underbrace{\mathrm{Im}(\rho_{B,R_{3}}\rho_{R_{3},M} e^{j 2\pi f(\tau_{B,R_{3}} + \tau_{R_{3},M})} \Lambda_{3}(\hat{\boldsymbol{\omega}}_{3}) \cdot x \cdot \Xi_{3})}_{\theta_{Imag}} + \mathrm{Im}(n).
\end{split}
\end{equation}
If we define $ p (y;\, \theta) $ as the likelihood function of the random variable $ y $ conditioned on $ \theta $, we have:
\begin{equation}
\label{deqn_ex34}
p (y;\, \theta_{Real}) = \dfrac{1}{\sqrt{2\pi\sigma^2}} e^{\dfrac{-(y-\theta_{Real})^2}{\sigma^2}}.
\end{equation}
Due to the maximum likelihood estimation: $ \argmax_{\theta_{Real}} p (y;\, \theta_{Real}) = y $. If we repeat this procedure for $ \theta_{Imag} $, we have:
\begin{equation}
\label{deqn_ex35}
\hat{\theta}_{Real} = \mathrm{Re}(y_{3}(\hat{\boldsymbol{\omega}}_{3})), \quad
\hat{\theta}_{Imag} = \mathrm{Im}(y_{3}(\hat{\boldsymbol{\omega}}_{3})),
\end{equation}
where $ \hat{\theta}_{Real} $ and $\hat{\theta}_{Imag}$ are the estimation of $ \theta_{Real} $ and $ \theta_{Imag} $ respectively. In this regard, when we consider the noise, we use the estimation of signals without the noise, which are received signals based on the maximum likelihood estimation, so the obtained results also work in this case. 
\vspace{-0.25cm}
\section{Numerical Results}
\begin{figure}[h]
 	\centering
 	\includegraphics[width=7.8cm]{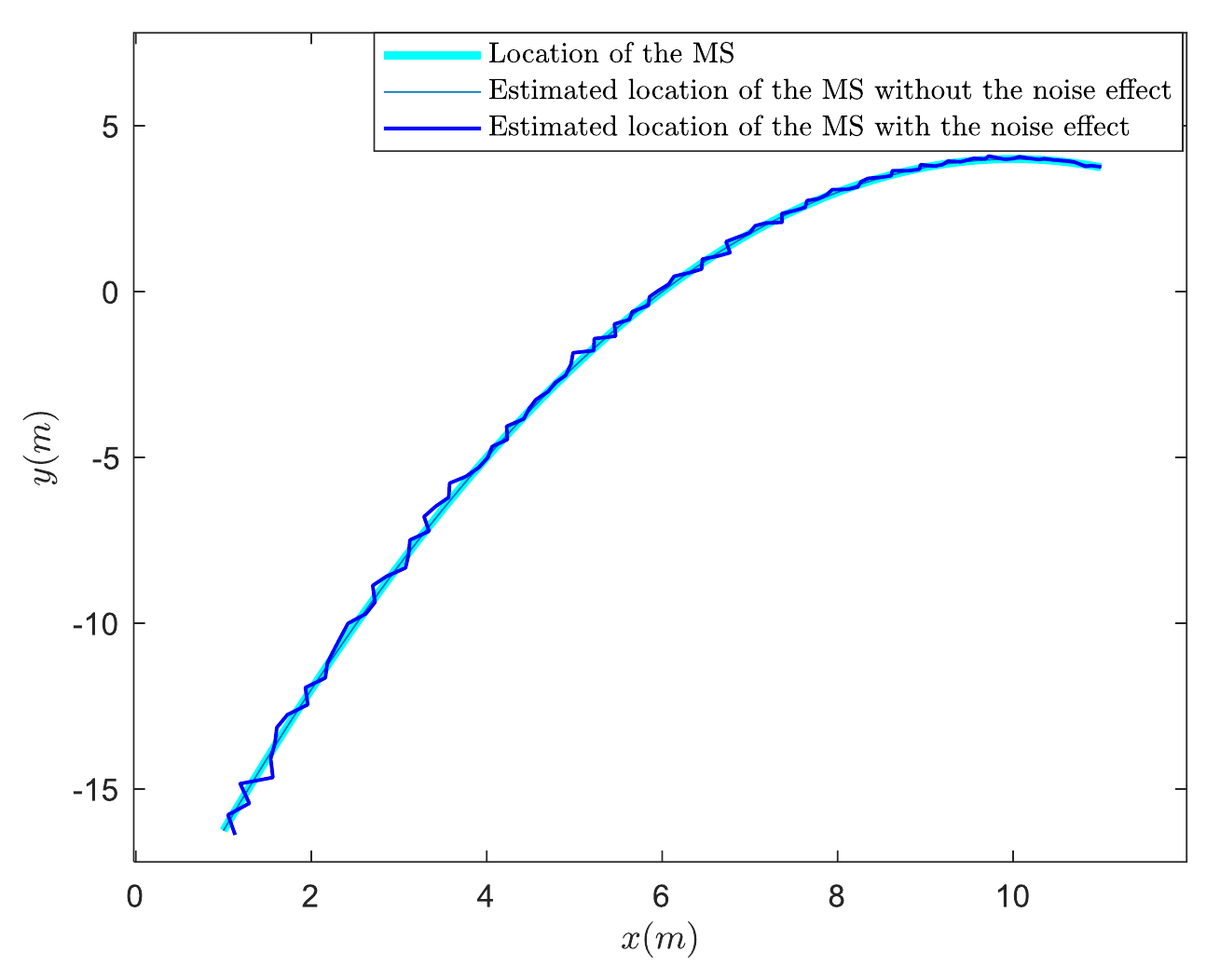}
 	\caption{Comparing the actual location of the MS, the estimated location of the MS without the noise effect, and the estimated location of the MS with the noise effect.}
 	\label{fig1}
 \end{figure}

\begin{figure}[h]
 	\centering
 	\includegraphics[width=7.8cm]{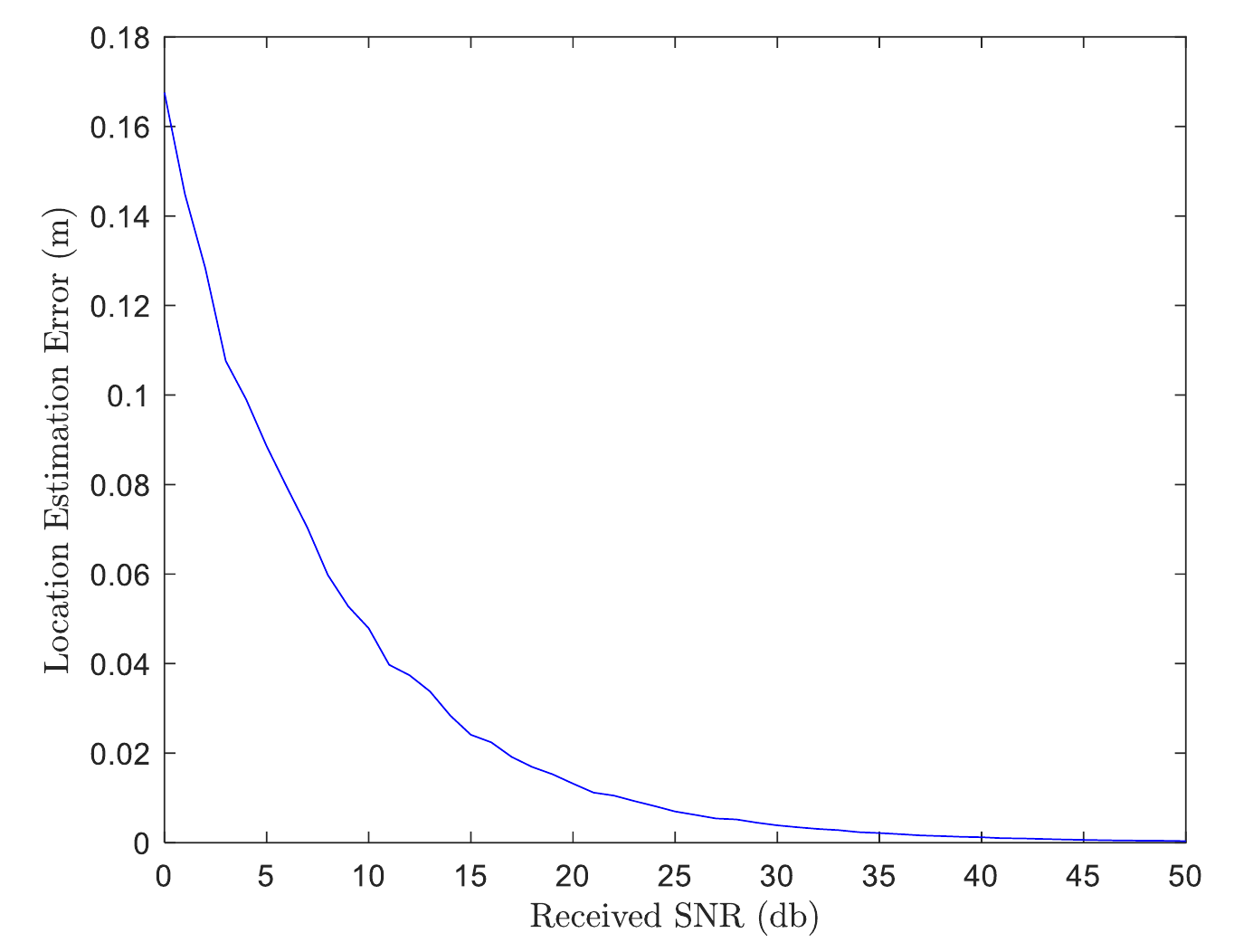}
 	\caption{The location estimation error versus the received $ \mathrm{SNR} $.}
 	\label{fig2}
 \end{figure}

\begin{figure}[h]
 	\centering
 	\includegraphics[width=7.8cm]{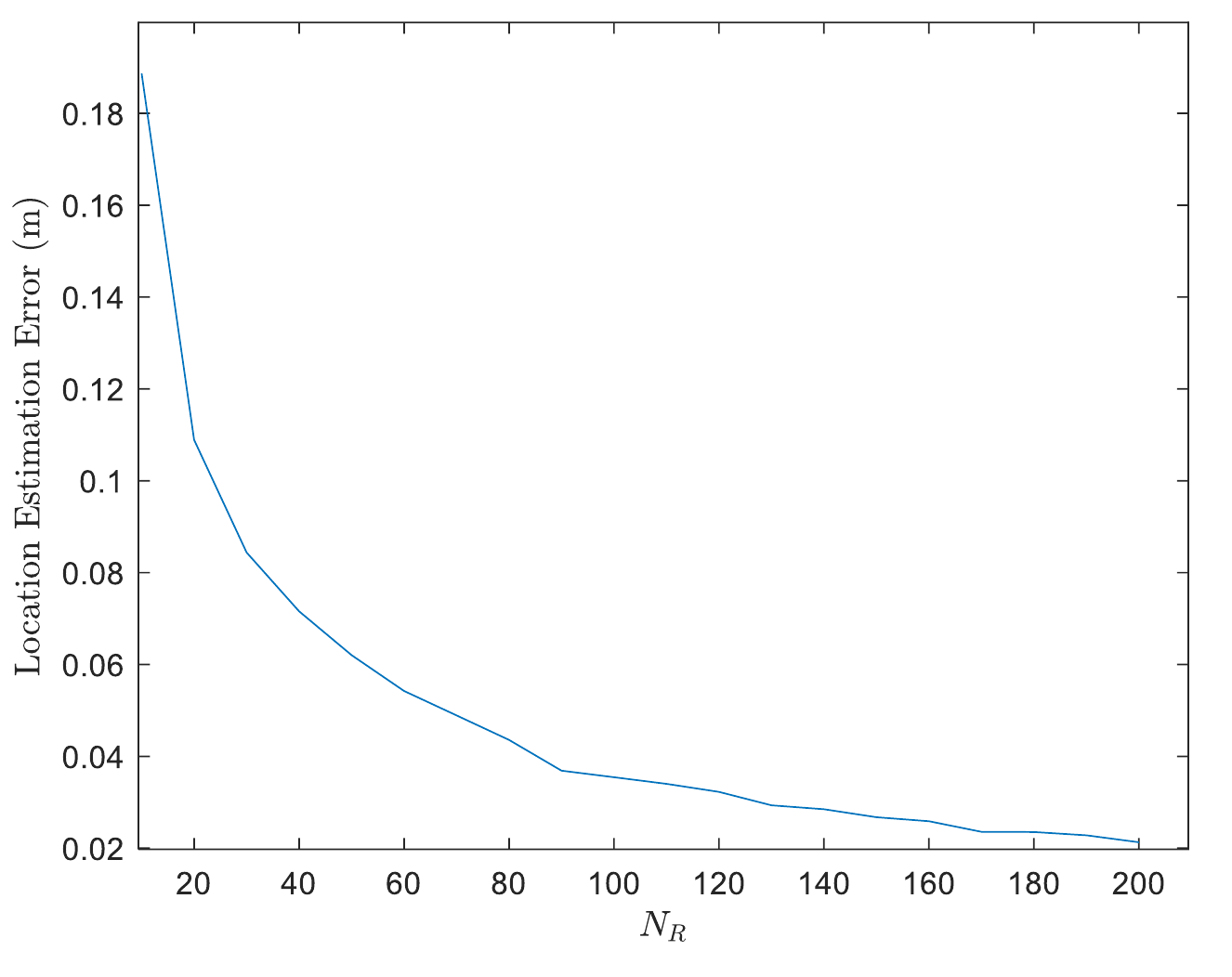}
 	\caption{The location estimation error versus the number of elements for each RIS.}
 	\label{fig4}
 \end{figure}

\begin{figure}[h]
 	\centering
 	\includegraphics[width=7.8cm]{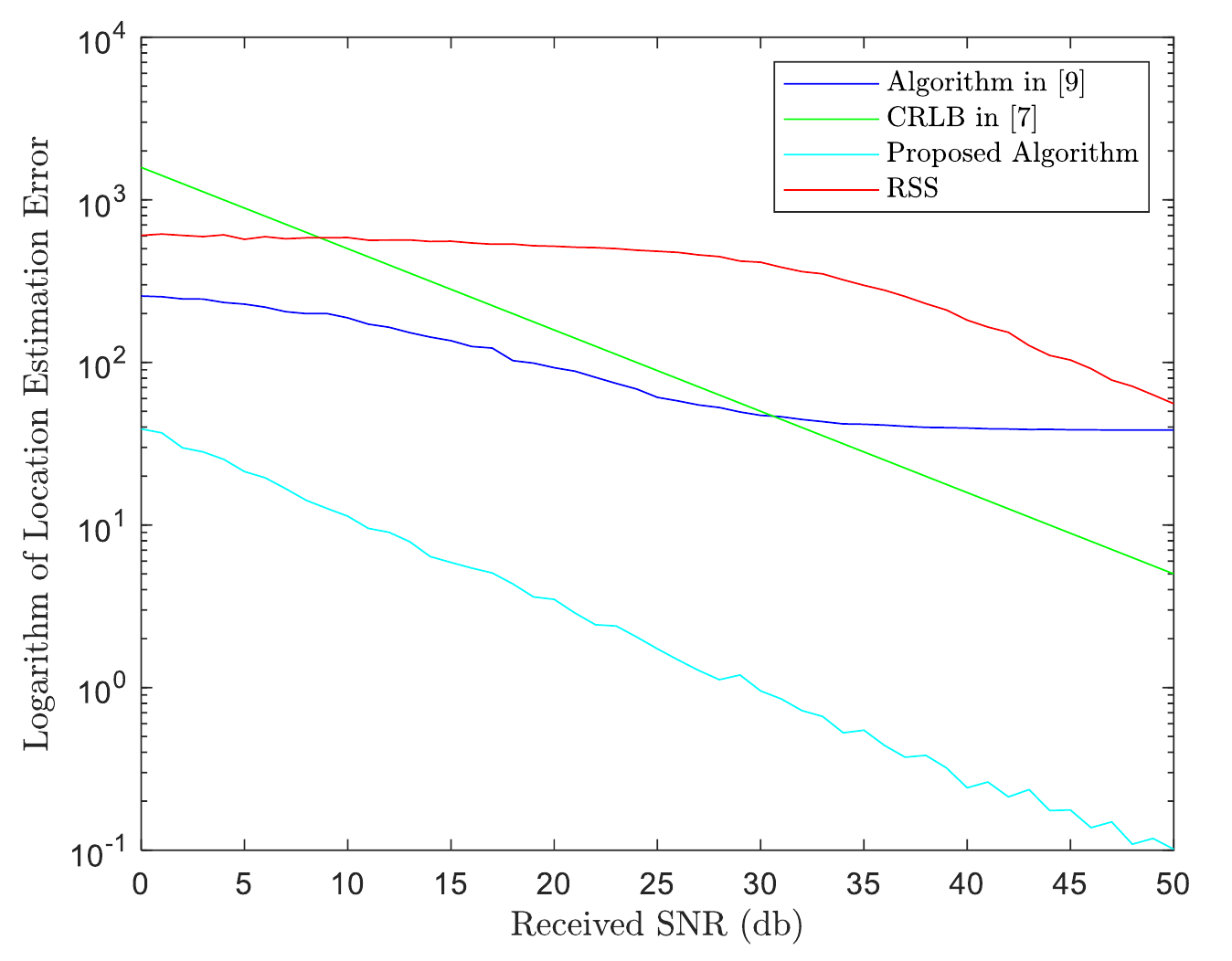}
 	\caption{Comparison of the location estimation error versus the received $ \mathrm{SNR} $ in the proposed algorithm, the algorithm in~\cite{nasri}, and the Cramer-Rao lower bound obtained for one RIS in~\cite{linear}.}
 	\label{fig3}
 \end{figure}

In this section, we use some simulations in order to illustrate the accuracy of the proposed algorithm. The position of the BS, RISs, and the MS are shown by $ B = (B_x, B_y, B_z) $, $ R_i = (R_{ix}, R_{iy}, R_{iz}) $, and $ M = (M_x, M_y) $, respectively. We set the parameters as $ B = (0, 0, 10) $, $ R_1 = (30, 20, 20) $, $ R_2 = (20, 40, 20) $, $ R_3 = (40, 40, 20) $, $ \mu = 2 $, $ d = \frac{\lambda}{4} = 3.75cm $, $ f = 2GHz $, $ N_{R_1} = N_{R_2} = N_{R_3} = 100 $, $ N_B = 20 $, $ \phi_{B,R_1} = \pi/6 $,  $ \phi_{B,R_2} = \pi/3 $, $ \phi_{B,R_3} = \pi/4 $, $ \theta_{B, R_1} = \pi/6 $, $ \theta_{B, R_2} = \pi/3 $, $ \theta_{B, R_3} = \pi/4 $, $ \theta_{R_1, M} = \pi/6 $, $ \theta_{R_2, M} = \pi/3 $, $ \theta_{R_3, M} = \pi/4 $. Note that the minimum distance between RISs and the BS is 20m, which is more than 100 times $ \lambda = 15cm $, so the BS and the MS are in the far field of the RISs, which is compatible with the system model.\\
\indent
The actual location of the MS, the estimated location of the MS when the noise is equal to zero, and the estimated location of the MS with the noise effect when the received $ \mathrm{SNR} = 12 \mathrm{dB} $ have been plotted in Fig. \ref{fig1}. It demonstrates that the estimated location error is zero when the noise is equal to zero. With the noise effect, however, the error exists, but the estimated location curve is a close approximation of the location curve without the noise.\\
\indent
Fig. \ref{fig2} depicts the error of the location estimation versus the received $ \mathrm{SNR} $ when the location of the MS is $ M = (60, 20) $. As expected, when the received $ \mathrm{SNR} $ increases, the error approaches zero. We further evaluate the location estimation error versus the number of elements for each RIS when the received $ \mathrm{SNR} = 12 \mathrm{dB} $. As it is depicted in Fig. \ref{fig4}, the location estimation error approaches zero as the number of elements in each RIS increases. \\
\indent
To compare the accuracy of the proposed algorithm with the accuracy of the algorithm in~\cite{nasri}, we set the parameters as $ B = (0, 0, 10) $, $ R_1 = (300, 0, 20) $, $ R_2 = (300, 300, 20) $, $ R_3 = (300, -300, 20) $, $ \mu = 2 $, $ d = \frac{\lambda}{150} = 0.001cm $, $ f = 2GHz $, $ N_{R_1} = N_{R_2} = N_{R_3} = 10 $, and $ N_B = 20 $. Note that we set $ d = \frac{\lambda}{150} = 0.001cm $ since for larger $ d $, the location estimation error of the algorithm in~\cite{nasri} not estimating the AoD is very large and is not comparable with the error of the proposed algorithm. In order to simulate the system model in~\cite{nasri}, we set $ N_B = 1 $ and use no estimation for AoD (the main differences between the proposed algorithm and the algorithm in~\cite{nasri}). Furthermore, we compare our algorithm with the Cramer-Rao lower bound for the scenario with one RIS obtained in~\cite{linear}, and RSS algorithm in~\cite{TOA}.
The location estimation error versus the received $ \mathrm{SNR} $ is shown in Fig. \ref{fig3} for our algorithm, the algorithm in~\cite{nasri}, the RSS algorithm in~\cite{TOA}, and the Cramer-Rao lower bound obtained for one RIS in~\cite{linear}. As can be seen, the accuracy of the proposed algorithm is much better than the others since the base station is multi-antenna, there are tree RISs, and AoD is estimated.

 \vspace{-0.25cm}
\section{Conclusion}
We proposed an algorithm to determine the position of an MS in a RIS-assisted environment. We illustrated that wireless localization is feasible when a BS and three RISs are available, and LOS is obstructed. We used a novel technique to eliminate the destructive effect of the AoD from the RISs to the MS.
Moreover, we calculated the location estimation error versus the received $ \mathrm{SNR} $ and versus the number of elements in each RIS in numerical results, and as expected, in the both cases the error approaches zero as the received $ \mathrm{SNR} $ and the number of elements increase. In this letter, we considered a scenario with obstructed LOS.
The direction of our future research is to generalize our proposed algorithm to scenarios, in which there exist LOS links and MSs are multi-antenna.

\bibliography{Paper_modified/refs}

\begin{thebibliography}{10}

\bibitem{RIS2}
E.~Basar, ``Transmission through large intelligent surfaces: A new frontier in wireless communications,'' in {\em 2019 European Conference on Networks and Communications (EuCNC)}, June 2019, pp. 112-117.

\bibitem{communication}
S.~Hu, F.~Rusek, and O.~Edfors, ``Beyond massive mimo: The potential of data transmission with large intelligent surfaces,'' {\em IEEE Trans. Signal Process.}, vol.~66, no.~10, pp.~2746--2758, May 2018.

\bibitem{localization}
H.~Wymeersch and B.~Denis, ``Beyond 5g wireless localization with reconfigurable intelligent surfaces,'' in {\em ICC 2020 - 2020 IEEE International Conference on Communications (ICC)}, June 2020, pp. 1--6.

\bibitem{passiveOrActive}
Z.~Zhang, L.~Dai, X.~Chen, C.~Liu, F.~Yang, R.~Schober, and H.~V. Poor, ``Active ris vs. passive ris: Which will prevail in 6g?,'' 2021. [Online]. Available: https://arxiv.org/abs/2103.15154.

\bibitem{TOA}
H.~Liu, H.~Darabi, P.~Banerjee, and J.~Liu, ``Survey of wireless indoor positioning techniques and systems,'' {\em IEEE Trans. Syst. Man Cybern., Part C (Applications and Reviews)}, vol.~37, no.~6, pp.~1067--1080, Nov. 2007.

\bibitem{RISLocalization}
A.~Elzanaty, A.~Guerra, F.~Guidi, and M.-S. Alouini, ``Reconfigurable intelligent surfaces for localization: Position and orientation error bounds,'' {\em IEEE Trans. Signal Process.}, vol.~69, pp.~5386--5402, Aug. 2021.

\bibitem{linear}
J.~He, H.~Wymeersch, L.~Kong, O.~Silvén, and M.~Juntti, ``Large intelligent surface for positioning in millimeter wave mimo systems,'' in {\em 2020 IEEE 91st Vehicular Technology Conference (VTC2020-Spring)}, May 2020, pp. 1--5.

\bibitem{obstructed}
Y.~Liu, E.~Liu, R.~Wang, and Y.~Geng, ``Reconfigurable intelligent surface aided wireless localization,'' in {\em ICC 2021 - IEEE International Conference on Communications}, June 2021, pp. 1--6.

\bibitem{nasri}
A.~Nasri, A.~H.~A. Bafghi, and M.~Nasiri-Kenari, ``Wireless localization in the presence of intelligent reflecting surface,'' {\em IEEE Wireless Commun. Lett.}, pp.~1315--1319, Apr. 2022.

\bibitem{future}
H.~Wymeersch, J.~He, B.~Denis, A.~Clemente, and M.~Juntti, ``Radio localization and mapping with reconfigurable intelligent surfaces: Challenges, opportunities, and research directions,'' {\em IEEE Veh. Technol. Mag.}, vol.~15, no.~4, pp.~52--61, Dec. 2020.

\end{thebibliography}

\bibliographystyle{ieeetr}
\end{document}